\documentclass[a4paper]{jpconf}
\usepackage{graphicx}
\newcommand{\Msun}{\ifmmode {M_{\odot}}\else${M_{\odot}}$\fi}

\begin{document}
\title{Constraints on physics of neutron stars from X-ray observations}

\author{Craig O. Heinke}

\address{Dept. of Physics, U. of Alberta, CCIS 4-183, Edmonton, Alberta T6G 2E1, Canada}

\ead{heinke@ualberta.ca}

\begin{abstract}
I summarize some constraints on the physics of neutron stars arising from X-ray observations of the surfaces of neutron stars, focusing on using models of  low-magnetic-field neutron star atmospheres to interpret their X-ray spectra.  I discuss observations of spectral lines, pulsation profiles, X-ray bursts, radius measurements of transiently accreting neutron stars in quiescence, crust and core cooling measurements of transiently accreting neutron stars, and cooling of young neutron stars.  These observations have constrained the neutron star mass and radius (and thus the internal composition, and dense matter equation of state), the superfluidity and neutrino emissivity properties of the core, and the composition and superfluid state of the crust.
\end{abstract}

\section{Introduction}

The behavior of matter at high densities (several times nuclear density) and relatively low temperatures ($kT << mc^2$) cannot be probed in terrestrial experiments.  The nonlinear aspects of quantum chromodynamics ensure that the equation of state cannot be extrapolated unambiguously from well-understood lower-density matter.  The interiors of neutron stars (NSs) are composed of high-density, relatively cold matter, and the physics of these interiors affect the observable properties (e.g. mass, radius, surface temperature) of the NSs.  Therefore, astrophysical observations of NSs permit constraints on the physics of dense matter, as have been summarized in detailed reviews such as \cite{Lattimer07,Ozel12}; see also Lattimer (in these proceedings) for a complementary discussion. 

NSs can be observed in a variety of manifestations \cite{Kaspi10}, many of which offer  possibilities for constraints on their internal physics.  In this short review, I focus on a few recent advances involving thermal X-ray radiation from the surface of the NS \cite{Pavlov03,Kaspi06}. Other constraints arise from, for instance, timing of radio pulses  \cite{Lyne90,Stairs04} which can provide measurements of the spin rate \cite{Hessels06} of NSs; neutrino emission from the formation of NSs in supernovae \cite{Kotake06}; gravitational waves as two NSs spiral together and coalesce \cite{Read09}; high-energy radiation produced by the decay of extremely high magnetic fields in the enigmatic NSs known as magnetars \cite{Mereghetti08}; and studies of binary companion stars and the accretion disks \cite{Bhattacharyya10}.  A key result is NS mass measurements in binaries with a radio pulsar, often using tests from general relativity (GR), which suggest that the majority of NSs are around 1.4 solar masses (\Msun) \cite{Kiziltan10}, with some ranging as high as 1.97 \Msun \cite{Demorest10}.

Here I discuss NSs where we observe thermal radiation from the surface (specifically, those with low magnetic fields, e.g. $B<10^{11}$ G). As they cool, isolated NSs radiate heat from their surfaces which may have nonuniform temperature distributions, leading to flux variations with the NS rotation phase.  NSs accreting material from a companion star experience occasional thermonuclear fusion events on their surfaces, ``X-ray bursts'', which often provide sufficient radiation to briefly expand the NS's atmosphere.  Accreting NSs often experience periods of quiescence during which material piles up in an accretion disk rather than reaching the NS, thus allowing one to study NS surface emission.  

\subsection{ Thermal surface emission}

NSs are born extremely hot, and their extreme gravity and magnetic fields can accelerate particles that heat their surfaces as NSs become older.  Thus, many NSs have temperatures of order a million degrees over part or all of their surface, and thus glow in X-rays.  This X-ray radiation is observed by X-ray telescopes on orbiting satellites (as X-rays cannot penetrate Earth's atmosphere); the findings reported below have principally come from NASA's {\it RXTE} and {\it Chandra}, and ESA's {\it XMM-Newton} satellites.

The surface radiation from NSs must pass through the thin (few cm) NS atmospheres (if present), which modify the radiation through free-free absorption, spectral lines, and photoionization jumps. 
Strong magnetic fields heavily distort the structure of electronic orbitals, and thus alter the ionization energies and opacities \cite{Lai97}.  Thus, it is easier to get constraints on NS physics when studying those NSs which are expected to have relatively low surface magnetic fields ($B$$<$$10^{6}$ T), as the number of free parameters affecting the spectrum is reduced.  Accretion of material from a companion appears to reduce NS magnetic fields by diamagnetic screening, from typical birth fields of $10^8$ T to $\sim$$10^4$ T \cite{Cumming01}. 

Due to the strong surface gravity, NS atmospheres stratify by element within $<100$ s \cite{Romani87}.  Thus, at times of little or no accretion, the part of NS atmospheres that produce the spectra is usually assumed to be pure hydrogen (during X-ray bursts, the composition should be complex, and one young NS appears to have a carbon atmosphere \cite{Ho09}).  Low-B-field hydrogen atmosphere models give observable X-ray spectra similar to blackbodies, but shifted to higher energies due to the inverse dependence of the electron free-free opacity on photon energy \cite{Rajagopal96,Zavlin96}.  The intense gravity of NSs shifts the observed spectrum to longer wavelengths, decreasing the observed temperature.  Although they complicate the spectral interpretation, gravitational effects permit constraints on the compactness (mass divided by radius) of the NS.

\section{Constraints}
\subsection{Spectral lines}

An unambiguous identification of a clear spectral line due to an atomic transition from a NS surface would permit, by measuring how the energy of the line is decreased by escaping from the gravitational potential of the NS, a straightforward measurement of the compactness of the NS. 
 A tentative identification of narrow absorption lines in a high-resolution X-ray spectrum of X-ray bursts from one accreting NS (EXO 0748-676) suggested a gravitational redshift of z=0.35 \cite{Cottam02}, consistent with theoretical expectations \cite{Lattimer07}, and sparking many attempts to find similar lines in other accreting NSs.  Unfortunately, deeper observations of this NS with multiple X-ray telescopes have failed to confirm these lines \cite{Cottam08}.  Furthermore, the NS rotation period, later measured at 552 turns per second \cite{Galloway10b}, would broaden any lines produced on the NS surface \cite{Chang05,Lin10}, implying that the observed  absorption lines do not arise from the NS surface.  Searches for absorption lines in high-resolution spectra of other accreting NSs have also met with no success.

Other spectral absorption lines have been observed in isolated NSs, both in nearby NSs \cite{Haberl07,vanKerkwijk07} and in very young NSs \cite{Sanwal02,deLuca12}.  These spectral lines appear to all be consistent with variants of proton or electron cyclotron absorption (or emission) lines \cite{Suleimanov10a,Suleimanov10b}, thus their study can tell us the surface magnetic field strength but not the surface redshift of the NS.  Recently, a time-variable absorption line has been claimed in a normal pulsar, which shows a magnetic field strength incompatible with the line's production by cyclotron absorption \cite{Kargaltsev12}.  It is difficult to understand how a line produced by an atomic transition would have a variable energy.  However, the apparent lines could instead be due to an intrinsically complicated multiple-blackbody spectrum (cf. \cite{Gotthelf10}), which when fit by a single blackbody gives the appearance of absorption lines \cite{Kargaltsev12}.  

\begin{figure}[h]
\begin{minipage}{19pc}
\includegraphics[angle=270, width=19pc]{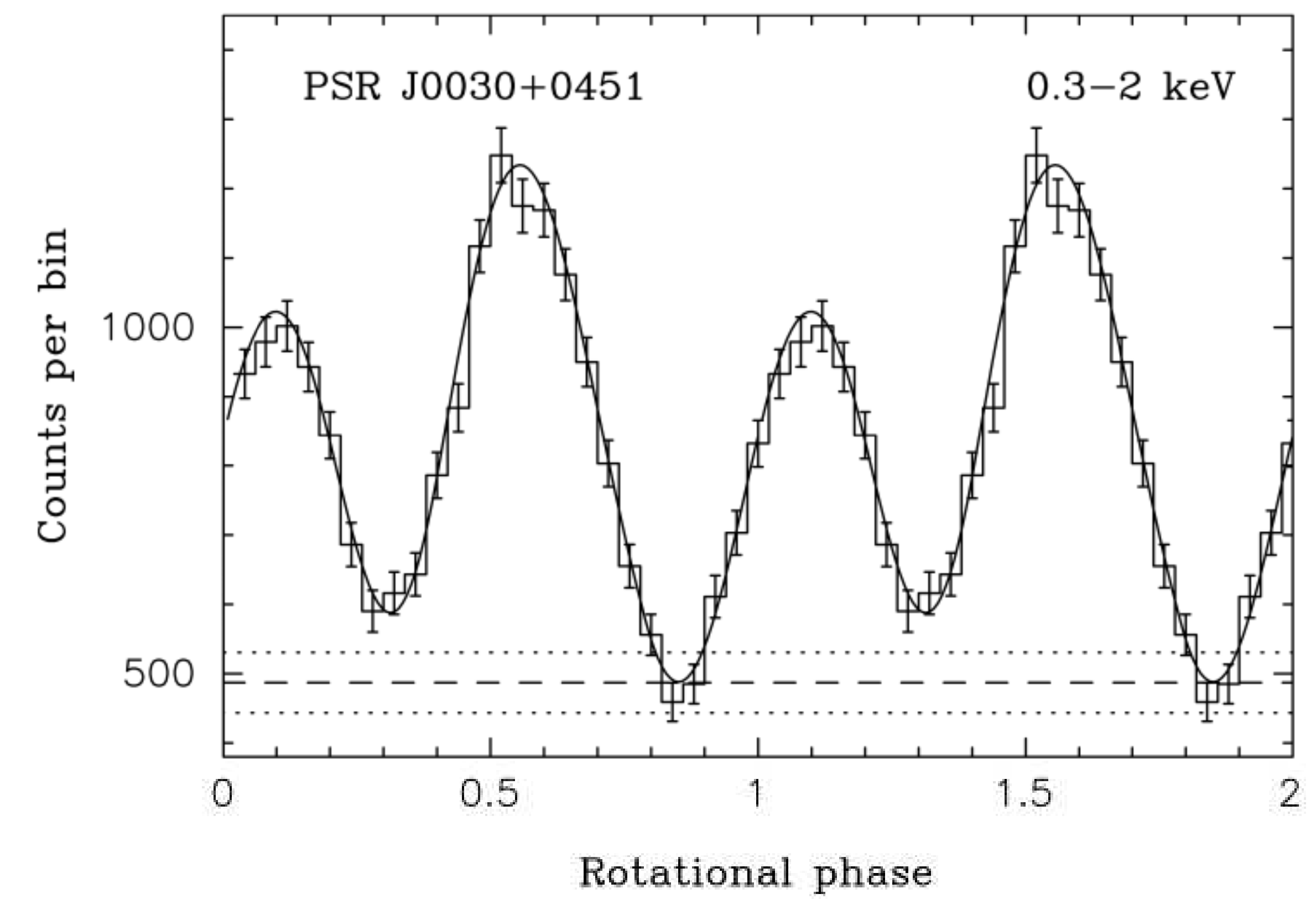}
\caption{\label{Bogd09} X-ray pulsation profile of the radio pulsar PSR J0030+0451 folded on the NS pulse period (repeated twice for clarity), as observed by the XMM-Newton telescope and modeled by \cite{Bogdanov09}.  The two peaks indicate the maximum visibilities of the two polar caps.}
\end{minipage}\hspace{1pc}%
\begin{minipage}{19pc}
\includegraphics[width=19pc]{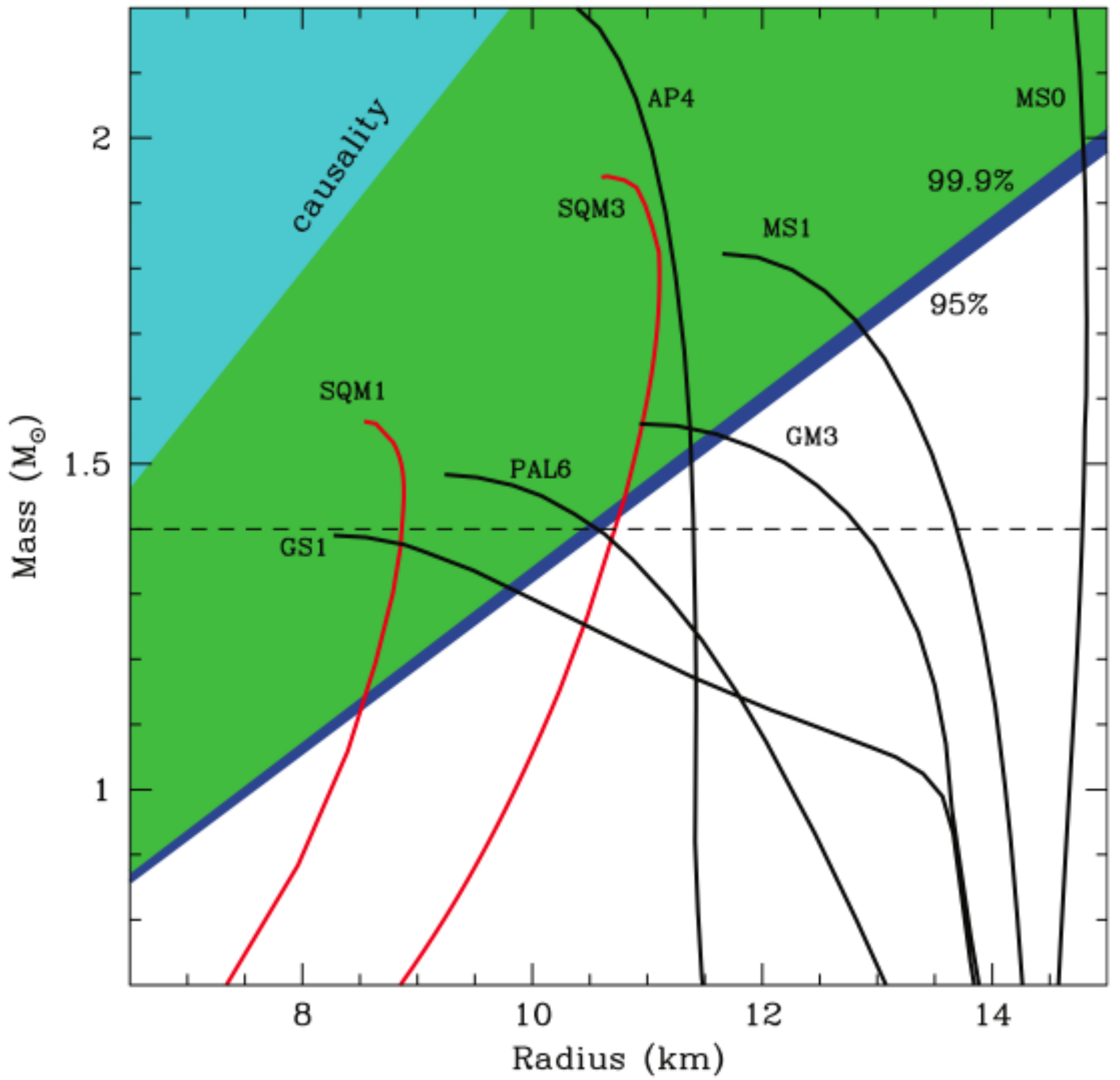}
\caption{\label{Bogd09b} Constraints on the mass and radius of the NS in PSR J0030+0451 from its pulsation profile \cite{Bogdanov09}.  The cyan region at upper left is not allowed due to causality (the requirement that pressure waves must propagate at speeds below $c$).  The region ruled out by fits to this NS is the green shaded region (at 99.9\% confidence); the dark blue region is also included at 95\% confidence. Several loci of mass and radius for different NS models are also plotted \cite{Lattimer01}.  The dashed line is the canonical NS mass, 1.4 \Msun.}
\end{minipage} 
\end{figure}

\subsection{Pulsation profiles}

Polar caps of NSs may be heated either by accretion, funneled onto the caps by the NS's magnetic field, or by radio pulsar activity, which produces energetic (relativistic) positrons and electrons that collide with and heat the polar caps \cite{Harding02}.  Either situation produces temperature inhomogeneities over the NS surface, which combine with the angular dependence of intensity from the atmosphere and ray-tracing in a general relativity metric to produce a complicated pulse profile \cite{Pechenick83}.  Generally, the Schwarzschild metric combined with appropriately oblate (due to their rotation) neutron stars, Doppler effects, and light-travel time delays provides a good approximation to the full GR lightcurve calculation \cite{Morsink07}.

Recent work has focused on accreting millisecond pulsars and radio millisecond pulsars, which are both believed to have relatively low magnetic fields.  The accreting millisecond pulsars, due to continuing accretion of a range of elements, may have spectra reasonably described by blackbodies, while radio millisecond pulsars, without continuing accretion, should typically have hydrogen atmospheres.  The simplest constraint from modeling the pulsation profiles is that a relatively sharp, peaked profile requires a larger star, since radiation from a more compact star suffers more gravitational bending and thus produces a smoother pulse profile.  
A strong constraint on the radius of the NS in the radio pulsar PSR J0030+0451 was calculated by \cite{Bogdanov09}, giving $R>10.4$ km (99.9\% confidence) for an assumed mass of 1.4 $M_{\odot}$, due to its highly peaked pulsation profile (Fig. 1).

More complicated modeling of the emission from accreting millisecond pulsars (which have components of their spectra that do not come from the surface, e.g. Comptonized radiation from above the surface, and reflected light) produces constraints in mass and radius space.  For the well-studied object SAX J1808.4-3658, \cite{Morsink11} find a relatively low mass below 1.5 $M_{\odot}$ (using a reasonable constraint on the inclination).  Although this modeling involves more parameters, the mass constraint is  
 consistent with optical studies of the companion star \cite{Elebert09}.

\begin{figure}[h]
\begin{minipage}{22pc}
\includegraphics[width=22pc]{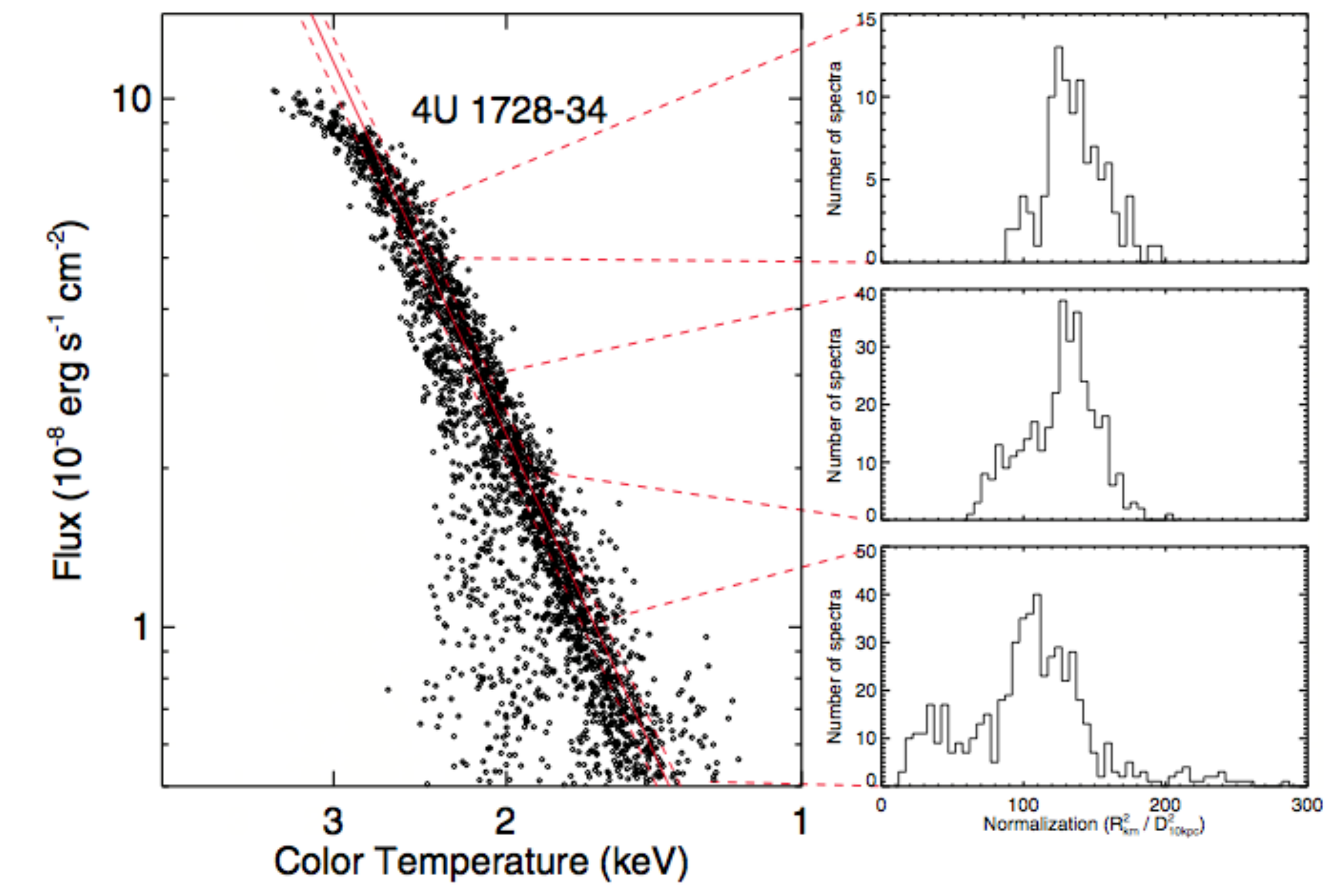}
\caption{\label{Guver12a} Flux vs. temperature for X-ray spectra taken at various times during multiple X-ray bursts by the same NS \cite{Guver12a}.  Diagonal lines indicate the best-fit blackbody normalization and its uncertainty. The majority of bursts are consistent with one normalization. }
\end{minipage}\hspace{1pc}%
\begin{minipage}{15pc}
\includegraphics[width=16pc]{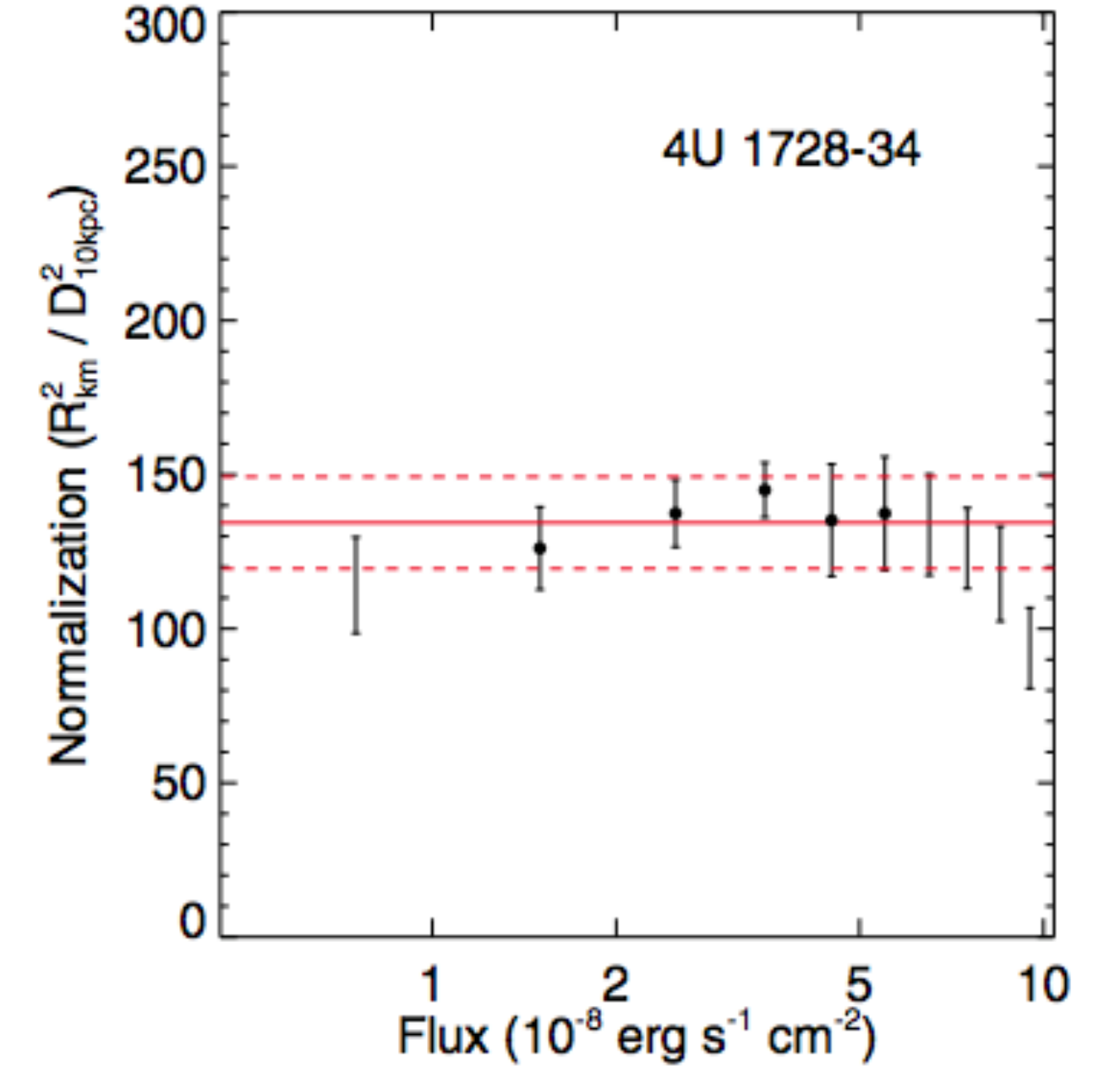}
\caption{\label{Guver12b}Inferred blackbody normalization vs. X-ray flux for the X-ray burst spectra of Fig. 3 \cite{Guver12a}.}
\end{minipage} 
\end{figure}

\subsection{X-ray bursts}

Accreting NSs often experience unstable nuclear burning on their surfaces, leading to rapid X-ray brightenings known as ``bursts'' \cite{Lewin95}.  Several authors have calculated models for NS atmospheres during bursts, showing how the measured temperature (defined by fits to a blackbody spectrum) relates to the effective surface temperature; this relation is parametrized as a ``color correction factor'' \cite{Majczyna05,Suleimanov11}.  The burst may expand the NS atmosphere; the radiative flux necessary to overcome gravity and push the atmosphere away, the ``Eddington limit'', depends on the NS mass, but the observed flux will be decreased by the NS's gravity, and thus its compactness.  As the radiation from the burst comes from the NS surface (rather than the physically complicated accretion flow that affects the radiation at other times), measurement of the temperature and flux of these bursts (and knowledge of their distances) determines the emitting area, and thus (after including GR corrections) the NS radius \cite{vanParadijs79}.  Similarly, measuring the radiation flux during bursts that expand the atmosphere tells us about the surface gravity of the NS \cite{Damen90}.  

Using time-resolved spectra of several bursts showing atmospheric expansion,  \cite{Ozel09,Guver10a,Guver10b} calculated constraints on the mass and radius of three NSs, finding relatively small NSs (with radii of 8-11 km).  These calculations have been criticized by \cite{Steiner10} on their choices of color correction factors, distances, and surface chemical compositions, and for not being internally self-consistent; the key problem is that the best-fit values of the observable properties produce complex (non-real-valued) values for the NS mass and radius.  \cite{Steiner10} present alternative calculations tending toward slightly larger ($\sim$12 km) NS radii, favoring the interpretation that the observed radius of the atmosphere remains larger than the true NS radius throughout the burst.  \cite{Suleimanov11b} use their color correction factors to derive a large radius ($>$14 km) for a burst from one NS, though they have also been criticized \cite{Guver12a} for the relatively poor fits of their spectra to blackbodies.

Several uncertainties surround the interpretation of X-ray bursts.  A critical question is whether the NS radii measured during the course of a burst, and between bursts, are identical and equal to the NS radius \cite{Galloway12,Guver12a}. Fig. 3 (from \cite{Guver12a}) shows temperature vs. flux at different times across multiple bursts from one NS, showing consistency of most burst spectra with one inferred radius (Fig. 4), which suggests that this uncertainty is manageable.  Other questions include whether the burst fluxes consistently reach the same Eddington limit \cite{Boutloukos10,Guver12b}, whether they show significant anisotropy  \cite{Zamfir12}, and whether they vary in their chemical composition \cite{Bhattacharyya10b}.  

\begin{figure}[h]
\includegraphics[width=22pc]{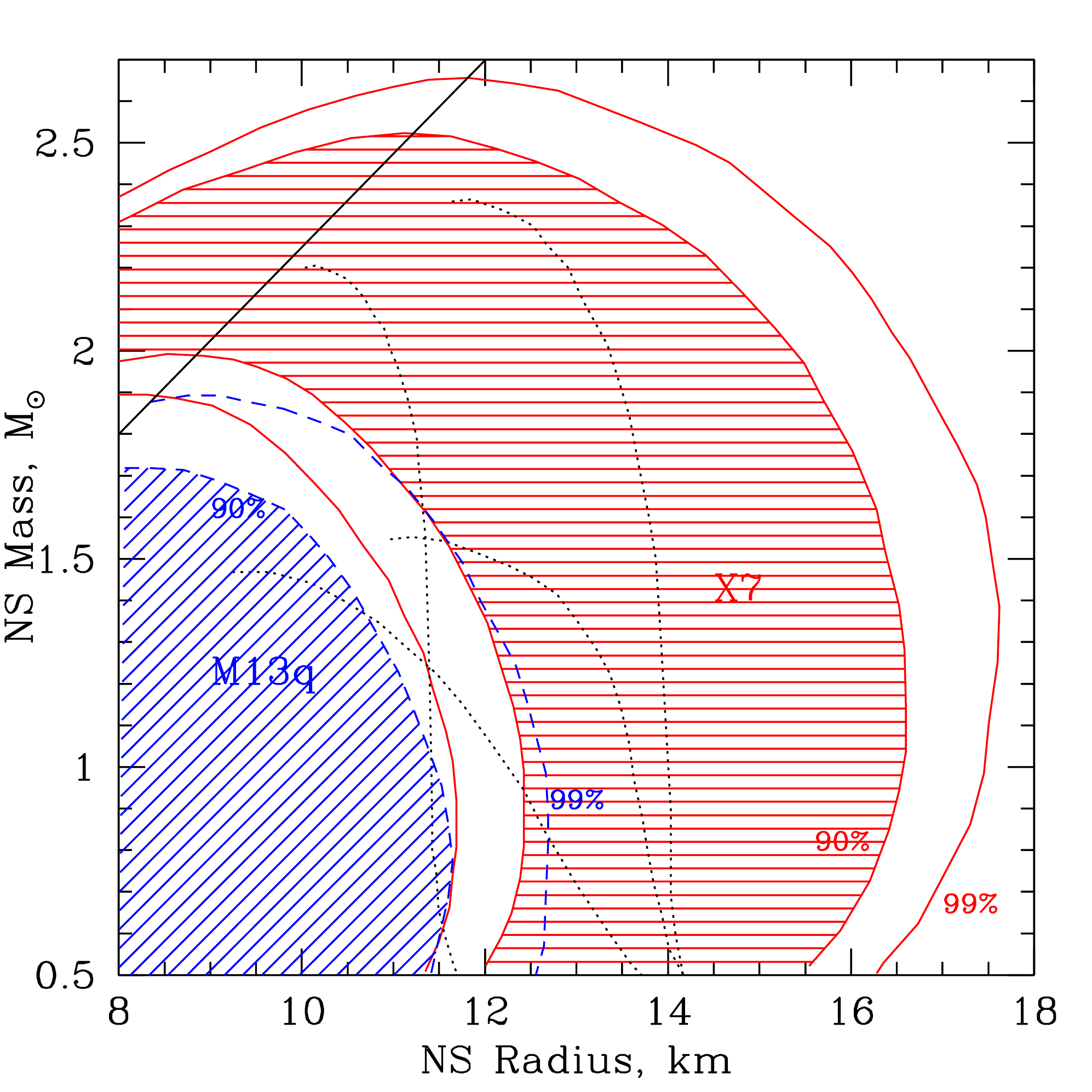}\hspace{1pc}%
\begin{minipage}[b]{14pc}\caption{\label{WebbvHeinke} Constraints on the mass and radius of two transiently accreting NSs at known distances, as inferred from fitting their X-ray spectra with hydrogen atmosphere models.  Hatched areas indicate the 90\% confidence regions for NSs in 47 Tuc (red) \cite{Heinke06a} and M13 (blue) \cite{Webb07}, with outside contours (solid for 47 Tuc, dashed for M13) marking the 99\% confidence regions of each.  Some sample NS equations of state are indicated by dotted black lines \cite{Lattimer01}.  The 90\% confidence contour regions for these two NSs do not overlap.}
\end{minipage}
\end{figure}

\subsection{Quiescent accreting neutron stars: radius}

Many accreting NSs show transient behavior, with periods of little or no accretion, in which the surface of the heated NS may be studied.  Assuming a low-magnetic-field hydrogen atmosphere without anisotropies, and a distance known by other means (e.g. by studying the other stars in its stellar cluster), tight constraints may be placed on the NS mass and radius by measuring the temperature and flux, and thus inferring the radius \cite{Rutledge02a}.  Due to the gravitational redshift, curved constraints in mass and radius are produced (e.g. Fig. 5).

Deep X-ray observations of globular clusters containing transient accreting NSs have placed tight constraints on the mass and radius of several NSs \cite{Heinke06a,Webb07,Guillot11a,Servillat12}.  Not all these constraints are consistent with each other (see Fig. 5), which may be due to instrumental effects \cite{Heinke06a}, and/or variations in atmospheric composition between NSs \cite{Servillat12}.  Combining results from these NSs with X-ray burst data (see above) gives robust constraints on the NS radius (between 10.4 and 12.9 km for 1.4 \Msun\ NSs), and thus on the dense matter  
equation of state \cite{Steiner12}.

\begin{figure}[h]
\includegraphics[width=24pc]{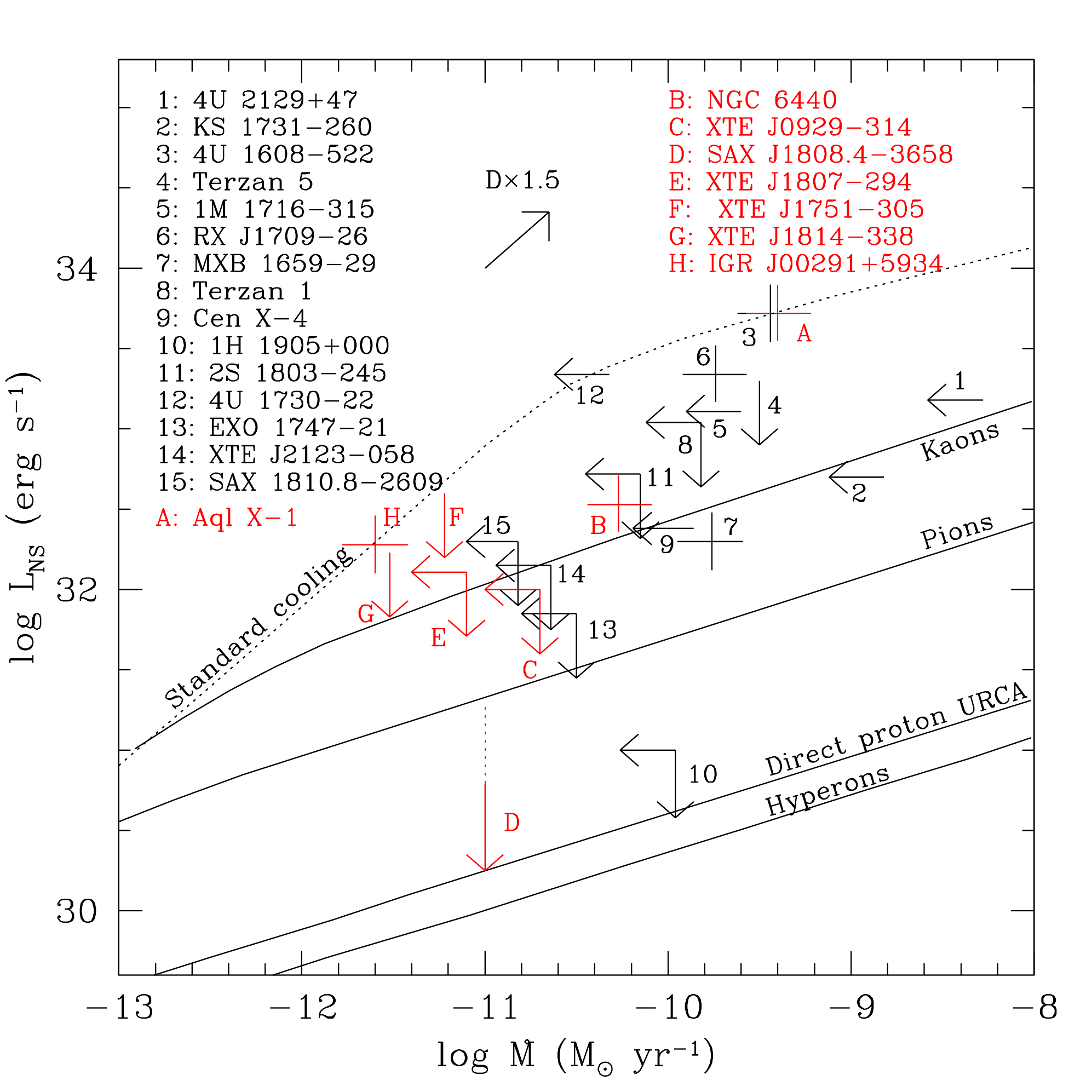}\hspace{1pc}%
\begin{minipage}[b]{14pc}\caption{\label{Heinke09} Cooling curves for various neutrino emission scenarios \cite{Yakovlev04}, compared with measurements or upper limits of the NS luminosity outside accretion, vs. the time-averaged mass transfer rate  \cite{Heinke09a}.  The dotted line relates heat input (via mass transfer) to the NS luminosity between accretion episodes for ``standard'' (slow) cooling, while the other diagonal lines suggest the maximal cooling achievable by specific enhanced cooling processes.  The effect of a factor of 1.5 error in distance is indicated.  SAX J1808-36's cold temperature (D) rules out most neutrino emission scenarios.}
\end{minipage}
\end{figure}

\subsection{Quiescent accreting neutron stars: core temperature}

Observations of transient NSs between periods of accretion measure the temperature that the NS has been heated to during accretion episodes.  The kinetic energy of the infalling material is radiated away immediately, but the added material compresses the deep crust, leading to nuclear reactions that deposit heat deep in the crust, some of which flows into the core \cite{Haensel90,Brown98}.  Averaged over long periods including many accretion episodes, the heat flowing into and out of the core will reach equilibrium, and the NS surface (between outbursts) will reflect the core temperature.  Neutrino emission from the NS core dominates the NS's energy loss, either through ``standard'' cooling (neutron-neutron neutrino bremsstrahlung processes) or enhanced neutrino cooling mechanisms such as direct URCA (involving protons or hyperons) or URCA-like processes involving (e.g.) pion or kaon-like Bose-Einstein condensations in the core, which will decrease the NS surface temperature \cite{Yakovlev04,Levenfish06}.  Differences in the surface temperature of NSs with the same average mass transfer rate (and thus heating history) can be explained by different NS masses, as this will change the central NS density and perhaps composition, and thus alter which cooling mechanisms may be accessed.

Strong evidence for enhanced neutrino cooling of the cores of transient NSs was provided by relatively cold NSs whose accretion histories suggested higher NS temperatures \cite{Colpi01,Wijnands01,Jonker06}.  Perhaps the clearest evidence for enhanced neutrino cooling is the cold NS SAX J1808.4-3658, where the mass transfer rate is very well-known (Fig. 6); this object may require direct URCA processes involving protons or hyperons \cite{Heinke07a,Heinke09a}.  An important question is whether the observed accretion history accurately reflects the long-term averaged mass transfer rates (see \cite{diSalvo08} and \cite{Patruno12} for differing views on this NS).

\begin{figure}[h]
\includegraphics[width=24pc]{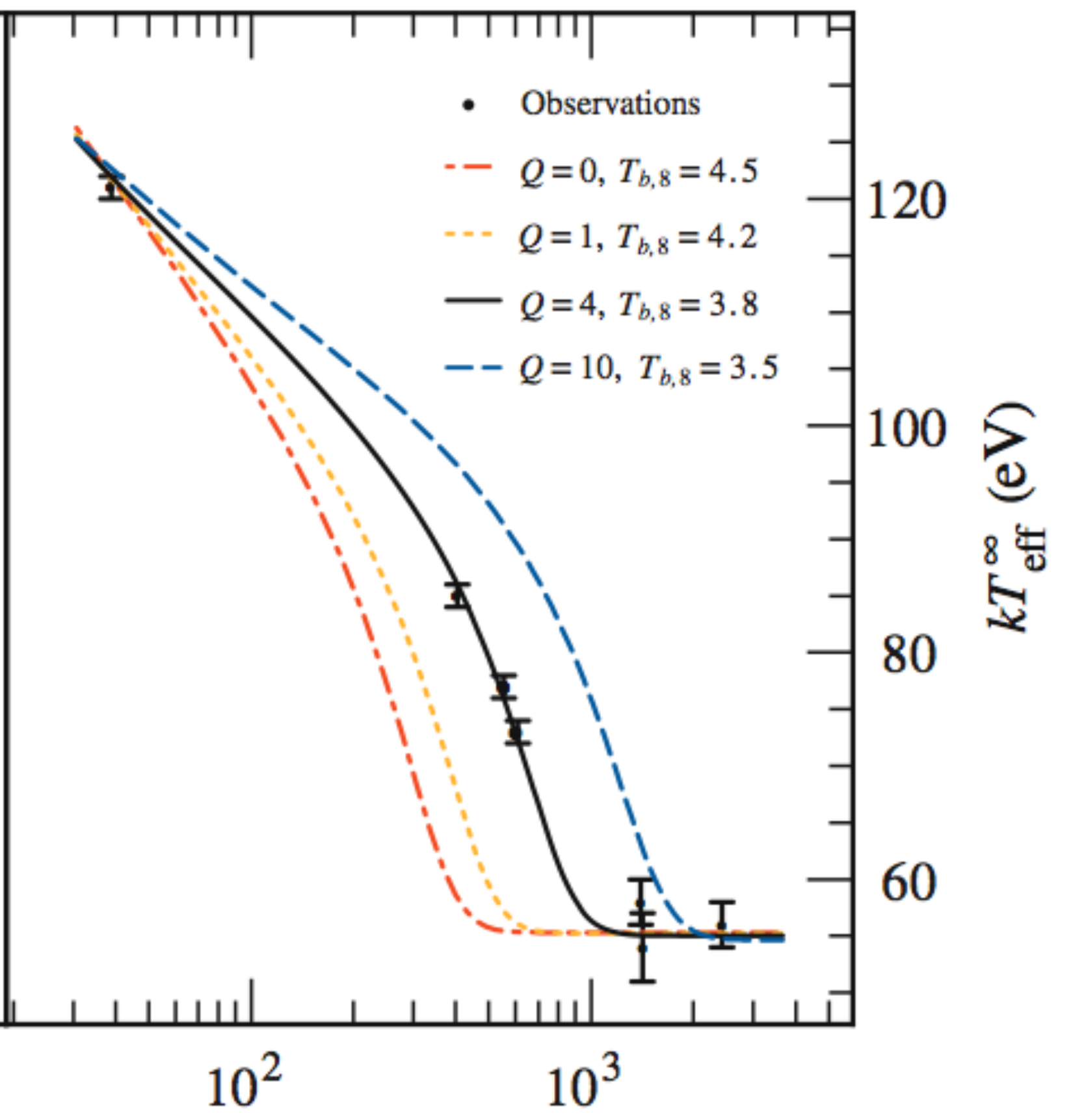}\hspace{0pc}%
\begin{minipage}[b]{14pc}\caption{\label{Brown09} Surface temperature vs. time for the transiently accreting NS MXB 1659-29 after accretion stopped, vs. models with different choices of crustal impurities (the initial temperature after accretion is adjusted to match the first point) \cite{Brown09}.  X-axis is days since accretion stopped, y-axis is NS surface temperature. The level of impurities (elements that disrupt the crystalline structure) in the crust is tightly constrained.}
\end{minipage}
\end{figure}

\subsection{Quiescent accreting neutron stars: crust temperature}

Observations of transient NSs immediately after long periods of accretion allow measurements of the cooling behavior of the crust, which radiates the heat produced by nuclear fusion within the crust within a few years, coming back to equilibrium with the core \cite{Rutledge02b}.  Crustal cooling has now been tracked for five transients over several years \cite{Cackett08,Cackett10,Fridriksson10,DiazTrigo11,Degenaar11a}.

The crustal cooling behavior is dominated by the crust's thermal conductivity, which is set by the level of defects in the crystalline crust, and its heat capacity, which is strongly reduced if the neutrons are superfluid.  Thus, measuring crustal cooling constrains the composition of the crust, and thus how nuclear reactions occur in the crust \cite{Horowitz09}.  Comparisons of the crust cooling curves with models (Fig. 7) show that the crust is thin, highly conductive (so crystalline with low impurities), and that the neutrons are indeed superfluid \cite{Shternin07,Brown09}.  A key remaining uncertainty is whether the measured NS surface temperature is affected by continuing, low-level accretion \cite{Fridriksson10}, which might appear to increase the rate of the temperature decay.

 \begin{figure}[h]
\includegraphics[width=24pc]{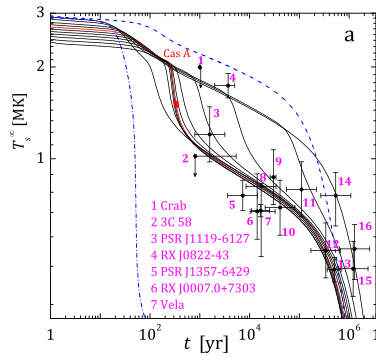}\hspace{2pc}%
\begin{minipage}[b]{14pc}\caption{\label{Shternin} Cooling curves (solid lines) for NSs with a range of masses (varying other parameters can reach the blue dashed lines), compared with observations (crosses) of the temperatures of young NSs of different ages.  The drop in model curves at around 100 years is due to the neutron superfluid transition.  The slope of the model curve passing through the Cas A NS data point (red) also matches the observed temperature drop of the Cas A NS \cite{Shternin11}. }
\end{minipage}
\end{figure}

 \subsection{Young cooling neutron stars}

The thermal behavior of NSs can also be studied by measuring the temperatures of young NSs of known ages, from 300 to roughly a million years old \cite{Yakovlev04}.  Here there is less uncertainty about the heat input and timescales (compared to transiently accreting NSs), but there is also a smaller range of temperatures.  Allowing for different compositions of the outer layers of the crust, \cite{Page06} showed that the known young NSs were consistent with a ``minimal'' (standard) cooling model, without the enhanced cooling mechanisms suggested to explain some of the transient NSs.  However, selection effects may affect this, as identifying young NSs often requires relatively high temperatures.  There are several supernova remnants which may contain young NSs, but sensitive, unsuccessful X-ray searches for them have constrained any such NSs to relatively low temperatures.  Considering the young ages of these supernova remnants, such cold NSs so early may require enhanced neutrino cooling \cite{Kaplan04}.

The youngest known NS, in the 330-year-old supernova remnant Cassiopeia A (Cas A), shows evidence of rapid cooling, by $\sim$3\% over 10 years \cite{Heinke10}.  It is too hot now to have experienced enhanced cooling, as those mechanisms would operate throughout its history and make it very cold already.  But its current temperature drop is too fast for normal cooling, requiring a recent transition to more rapid cooling.  This can be explained if the neutrons in the NS core are currently transitioning to a superfluid state; as the neutrons bind into the quantum pairs required to enter the superfluid state, they release the ``gap'' energy of the pairing in the form of neutrinos \cite{Flowers76}.  Models of the temperature decline from such a transition match the data, for a critical temperature of the neutron superfluid of $5-9\times10^8$ K \cite{Page11,Shternin11}; see Fig. 8.  The key question here is how well the calibration of the X-ray telescope (always a complex question) is known.  This is the first observationally driven estimate of the superfluid critical temperature for any superfluid other than the helium-3 and -4 isotopes on Earth, and a beautiful illustration of the potential NSs can offer for constraining properties of the superdense matter.

\ack
I thank my collaborators in the research I have done in these areas, particularly the theorists Wynn Ho, George Rybicki, Peter Shternin, Ron Taam, and Dima Yakovlev. I also thank S. Bogdanov, T. Guver, E. Brown, \& P. Shternin for permission to reproduce figures.  I am supported by an Ingenuity New Faculty Award and by an NSERC Discovery Grant.

\section*{References}

\bibliography{src_ref_list}
\bibliographystyle{vancouver}

\end{document}